# Direct imaging of thermally excited metastable structures of ion Coulomb clusters

Michael Drewsen*, Thierry Matthey†, Anders Mortensen* & Jan Petter Hansen‡

*QUANTOP – Danish National Research Foundation Center for Quantum Optics and, Department of Physics and Astronomy, Aarhus University, Ny Munkegade, 8000 Aarhus C, Denmark

Tel.: +45 8942 3752;   Fax: +45 8612 0740;   Email: drewsen@phys.au.dk

†Parallab, Bergen Center for Computational Science, University of Bergen, Thormøhlensgate 55, 5008 Bergen, Norway

‡Department of Physics and Technology, University of Bergen, Allegate 55, 5007 Bergen, Norway

**Coulomb crystallisation of large ensembles of ions has in the past years been intensively studied experimentally with many spectacular results of relevance to infinite systems in one-[1-4], two-[5], and three-dimensions[6-8]. While strings of a few ions have proven to be very attractive objects in quantum information processing[9,10], larger Coulomb crystals have very recently found applications within other aspects the dynamics of quantum systems[11-13]. Smaller finite ensembles of cold identical charged particles confined by a harmonic potential furthermore constitute very special types of clusters due to the pure repulsive long-range inter-particle forces[14-17]. Here, we report on the direct imaging of metastable structures of Coulomb clusters consisting of a few thousands confined and laser-cooled $^{40}Ca^+$ ions. The observations are attributed to structural excitations due to finite temperatures, a feature likely to appear in clusters of short-range interacting particles[18-22], but yet not observed directly.**

Generally, it has been found that Coulomb crystallisation occurs for identical charged particles when the criterion

$$\Gamma = E_{Coul}/E_{kin} > \sim 200, \qquad (1)$$

where $E_{Coul}$ is the average nearest-neighbour Coulomb energy and $E_{kin}$ is the kinetic energy of particle per degree of freedom, is fulfilled[14].

For larger Coulomb clusters (more than ~10.000 particles) confined by a harmonic potential, the ground state has both theoretically[16] and experimentally[6] been shown to be a body-centred-cubic (bcc) structure in contrast to a face-centred-cubic (fcc) structure for clusters where the constituencies interact via short-ranged centro-symmetric potentials[18] (e.g. Lennard-Jones or Morse potentials). For small Coulomb clusters, the specific ground state structure depends critically on both the confining potential and the number of particles[17]. This is a common feature for all types of clusters, since so-called magic numbers of particles exist due to filling of specific electronic or geometric shells[23]. However, as the number of particles increases the difference in the binding energy between various spatial configurations becomes smaller and at finite thermal energies the number of isomeric configurations is generally expected to grow exponentially with the number of particles[24]. Though it is predicted that at temperatures below the melting point various cluster systems will undergo a structural transition[18-20], only indirect experimental evidence exists so-far[21,22].

In this Letter we focus on the direct observation of thermally excited metastable structures in Coulomb clusters consisting of a few thousand particles. In Fig. 1a and 1b, examples of projection images (see Method) of nearly perfect spherical clusters of about 2700 $^{40}Ca^+$ ions are presented. While in Fig. 1a one clearly observes the fluorescence localized in ring-structures with an unordered core, indicating that the outer ions are localized in concentric shells, in Fig 1b a near-hexagonal structure is present. Since the depth of field of the imaging system is at least a few times the dimensions of the observed structures, the projection image must originate from a three-dimensional lattice structure[7]. Though simple cubic (sc), fcc and bcc structures all show up as the observed hexagonal structure when viewed along one of the cube diagonals (e.g., the [111]-direction) for a given side-length of the hexagonal structures, the three cubic structures form a hierarchy of densities $n$ such that $n_{bcc}=2n_{sc}=4n_{fcc}$. Only $n_{bcc}=(2.3\pm0.3)\cdot10^8$ cm$^{-3}$ (deduced from Fig. 1b) agrees with the expected density of $n=(2.2\pm0.1)\cdot10^8$ cm$^{-3}$ found from a simple relation between trap parameters and the ion density (see Method).

Since the observed bcc structure is in contradiction to MD simulations of the ground state of Coulomb clusters of this size[16], we have performed a series of MD simulations using ProtoMol[25] with a certain fraction of the ions in the core of the clusters kept fixed in a bcc-lattice with the expected density while the remaining ions were allowed to relax to minimize the energy of the entire ion system. More precisely, in the simulations, the free ions were first kept at a temperature around the melting point of the crystalline structure for 5 ms, after which the temperature during another period of 5 ms was linearly reduced to a temperature corresponding to $\Gamma>10.000$. At the end of each simulation the normalized cohesive energy per particle $u_{coh}=(U-U_{hom})/(Nq^2/a)$ was calculated with $U$ being the total potential energy of the ion system, $N$ the number of ions, $q$ the charge of the ions, $a$ the ion mean distance and $U_{hom}$ the potential energy of a homogeneously charged sphere with the same total charge and radius as the ion system. In Fig. 2, $u_{coh}$ is plotted as a function of the fraction of ions kept fixed. As expected, the cohesive energy is minimal when all ions are free to relax. However, keeping a smaller fraction of the ions fixed does not make the energy of the system increase dramatically. Since the ions in the experiments are cooled to a temperature corresponding to $\Gamma \sim 100 - 1000$, well below $\Gamma>10.000$ reached in the simulations, it is reasonable to find the ions in a configuration different from the one corresponding to minimum energy states in Fig. 2. In Fig. 3, we have plotted the cohesive energy of a system of 2685 ions as a function of temperature. The results have been found by MD simulations with all ions free to relax, and where the temperature to be reached was set to various values. At a typical temperature obtainable in the experiments of 5 mK ($\Gamma \sim 400$), we find that the cohesive energy is about 0.004 higher than the minimum energy of the same system in Fig. 2. This is, e.g., more than an order of magnitude larger than the energy excess of a system with 10 % of the ions fixed, indicating that systems with a bcc core are energetically allowed. In the insert of Fig.2, the projection image of the MD simulation with 10 % of the ions fixed looks very similar to the ones observed experimentally (see

Fig. 1b). To strengthen the point of view that the observed lattice structures are the consequence of thermal excitation, MD simulations have been performed where the initial configuration was the minimum energy state of a system having 10 % of the ions initially locked into a bcc structure, and where all the ions were subsequently slowly heated to a temperature of 5 mK and 20 mK, respectively. In Fig. 1c and 1d, the expected projection images of the ions, when averaging over a time window of 5 ms, are presented for the two temperatures. Clearly, for T=5 mK (Fig. 1d), the bcc-structure does not vanish on the 5 ms time scale, while for T=20 mK (Fig. 1c), only a shell structure with an unordered core is left to be seen. This is analogous to the two experimental cases shown in the same figure (Fig. 1a and 1b). These findings suggest that once an ordered state has been created, a barrier exits for relaxation towards the ground state as usually expected in cluster physics[26].

The bcc-structures are not the only thermally excited crystal structures to be observed, as has been demonstrated through a long term monitoring (several minutes) of the same cold ions in the trap. In fig. 4, a selection of images from a video sequence of a slightly prolate cluster of ~2700 ions is presented. Clearly, not only the previously discussed bcc-structure (Fig. 4a) is present. While Fig. 4b indicates smaller nucleation sites, Fig. 4c suggests that ions are organized partly in a fcc structure (<211>-direction of projection), and finally, Fig. 4d is consistent with the coexistence of a bcc and a fcc component joint at an interface between bcc (110) and fcc (111) planes in, e.g., a Nishiyama-Wasserman orientation[27]. For systems with up to 50 % ions fixed, MD simulations gave essentially the same binding energies for bcc, fcc and hcp structures. Hence, the lack of observation of the hcp structures might be a consequence of low energy barriers for relaxation towards the ground state or other configurations[26], or simply a result of the non-perfect cylindrical harmonic potential. The latter, have been documented to result in very specific and persistent structures in two-species Coulomb crystals[8].

The existence of metastable structures does not depend critically on the shape of the crystal or on the number of ions. They have been observed in crystals with aspect ratios of both ~1:2 and ~2:1 and with the number of ions ranging from ~1000 to ~10.000 (for the special case of bcc-structures, see Ref. 7)

As indicated by the images in Fig. 4, ion Coulomb systems offer a unique possibility of directly observing the dynamics of cluster structures, and hence constitute an interesting basis for investigations of thermodynamic and phase transitions of not only Coulomb clusters, but clusters in general. In addition, Coulomb clusters are extremely well suited for real-time studies of the gradual transition toward stable long-range ordered structures since the number of ions in the crystals can be controlled at the few-per cent level up to hundreds of thousands of ions without having to worry about impurities.

In contrast to chemically bounded clusters, the coupling strength can for ion Coulomb clusters easily be manipulated either globally by changing the confining trap potential or locally by introducing spatially tailored induced dipole potentials by utilizing off-resonance laser fields[28]. This gives potentially control of both the structures and their orientations. Furthermore, certain types of defects may be introduced by trapping more than a single ion species. Besides these fundamental cluster aspects, Coulomb clusters can be applied in a wealth of research fields ranging from quantum information[9-11] to cold molecular physics[29,30].

Methods

The experiments have been carried out on Doppler laser-cooled $^{40}$Ca$^+$ ions confined in a linear RF trap, the details of which can be found elsewhere[3,7]. By varying the RF voltage used to create the radial confinement or the static voltage creating the axial trapping potential both the ion density as well as the shape of the ion clusters can be controlled. The effective cylindrical symmetric harmonic trapping potential which may be written

$$\Phi(z,r) = 1/2\, m(\omega_z^2 z^2 + \omega_r^2 r^2), \quad r^2 = x^2 + y^2$$

leads at low temperatures to spheroidal shaped clusters with an average homogeneous density of

$$n_{ion} = m(\omega_z^2 + 2\omega_r^2)/(q\varepsilon_0),$$

with $m$ being the mass of the ions, $q$ being the charge of the ion and $\varepsilon_0$ being the vacuum electric permittivity.

The Coulomb clusters are imaged by collecting the fluorescence light emitted by the ions during the laser cooling process to a 2D CCD array. The imaging optic is situated such that the real crystal structure is projected to a plane including the z-axis. Due to the rotational symmetry of the clusters around the z-axis, we can from the images deduce their real 3D sizes and hence the number of ions.

**Figure 1** Projection images of Coulomb clusters. Images from experiments with clusters containing ~2700 ions (**a** and **b**). The blue colour-scale indicates the fluorescence level with light blue being high intensity. The fluorescence integration time was 0.1 s. Projection images based on data from MD simulations of Coulomb clusters with 2685 ions at a temperature of **c** 20 mK and **d** 5 mK, respectively. The presented scale bar corresponds to 150 μm.

**Figure 2** The cohesive energy per ion $u_{coh}$ found from MD simulations. The results originate from simulations of a Coulomb cluster with 2685 ions. The ordinate presents the ratios of ions fixed in a bcc structure. The insert presents expected observable projection images in experiments where no or 10 % of the ions are kept fixed, respectively.

**Figure 3** The cohesive energy per ion $u_{coh}$ as a function of the temperature. The results are obtained from MD simulations of a Coulomb cluster with 2685 ions. The dashed vertical red line indicates the temperature below which crystallisation is expected to appear ($\Gamma \sim 250$)[12].

**Figure 4** Projection images of prolate Coulomb clusters. Snapshots (0.1 s integration time) of a slightly prolate crystal of ~2700 ions. The color code is as in Fig. 1. **a**, Signature of a bcc structure viewed from the <111> direction. **b**, crystal structures with indications of nucleation sites. **c**, Crystal structure containing a fcc structure viewed from the <211> direction. **d**, Presumably an example of coexistence of a bcc and a fcc structure. The presented scale bar corresponds to 150 μm.

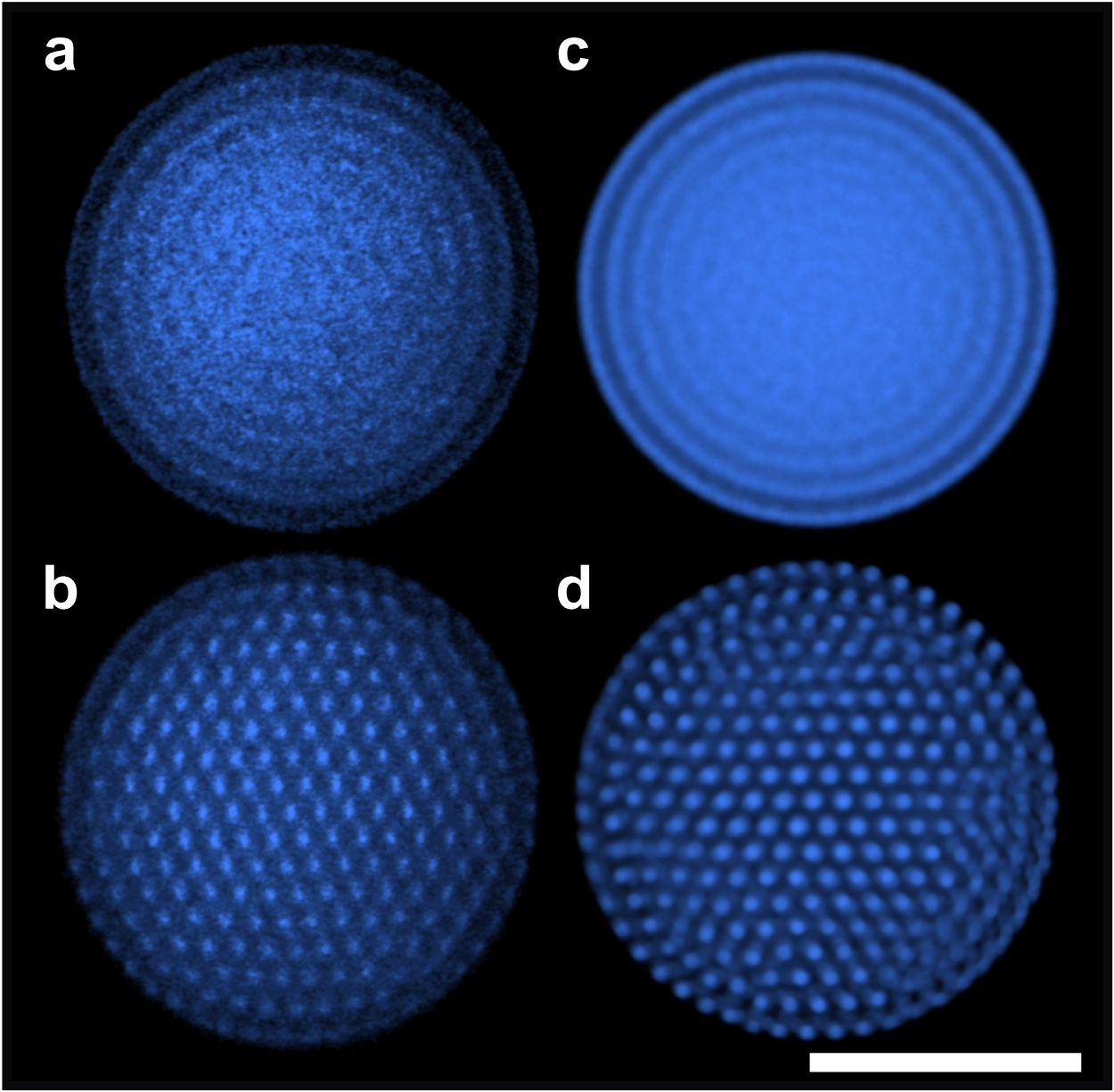

Drewsen, Fig. 1

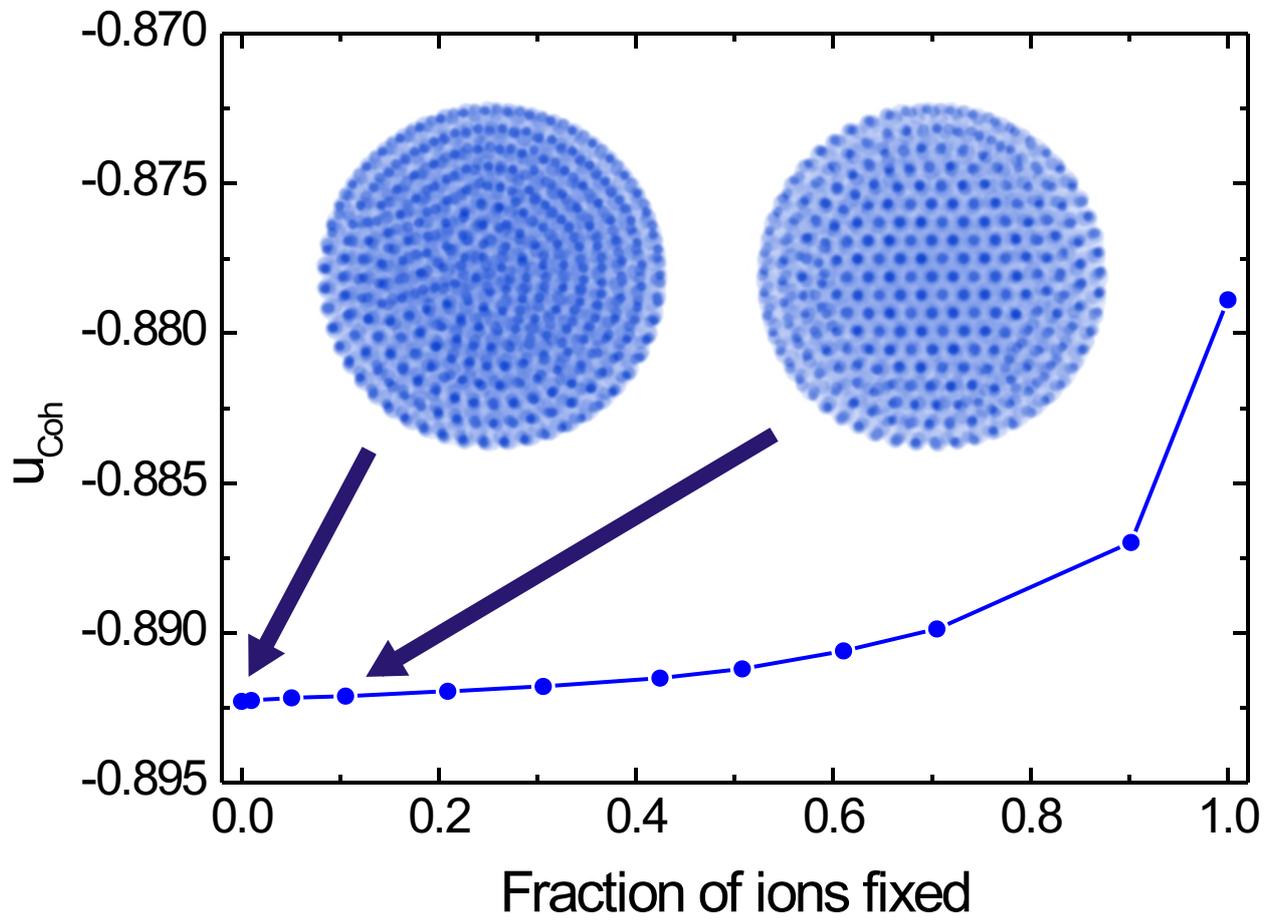

Drewsen, Fig. 2

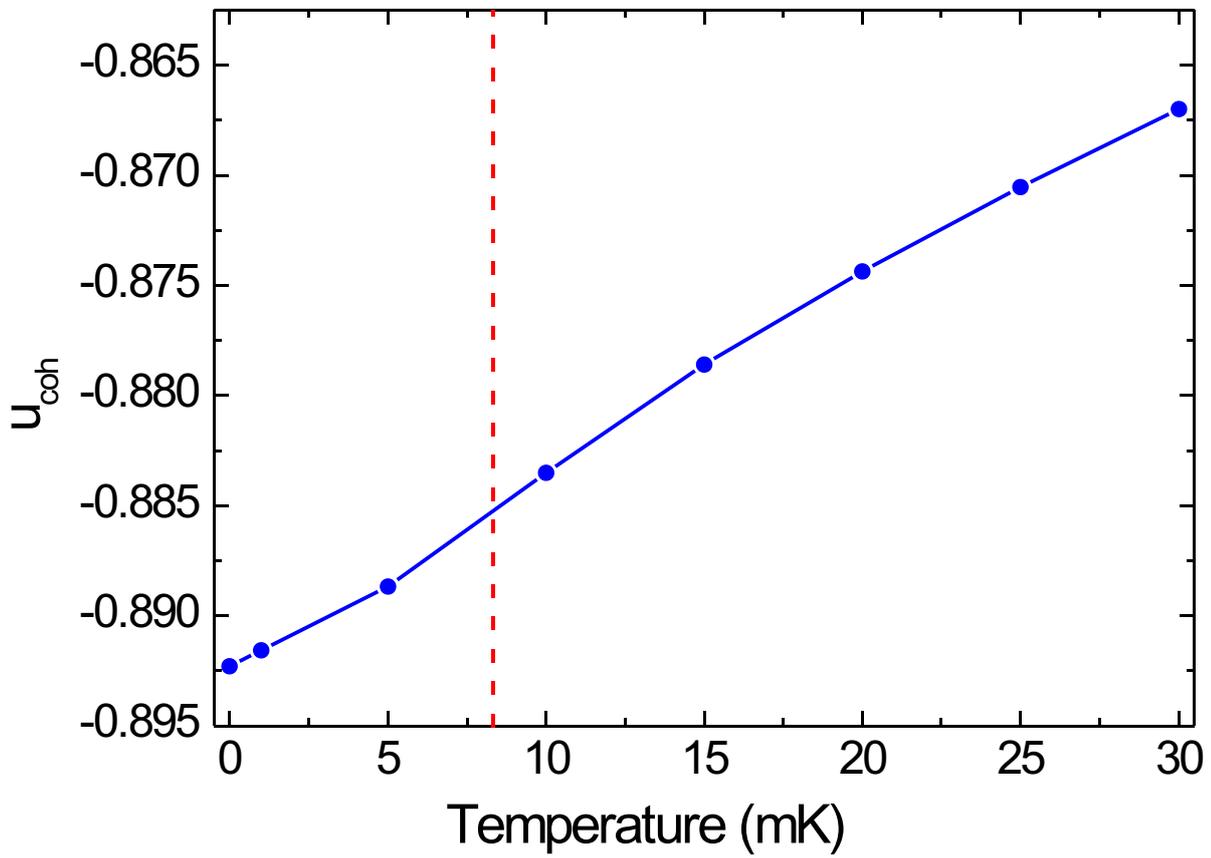

Drewsen, Fig. 3

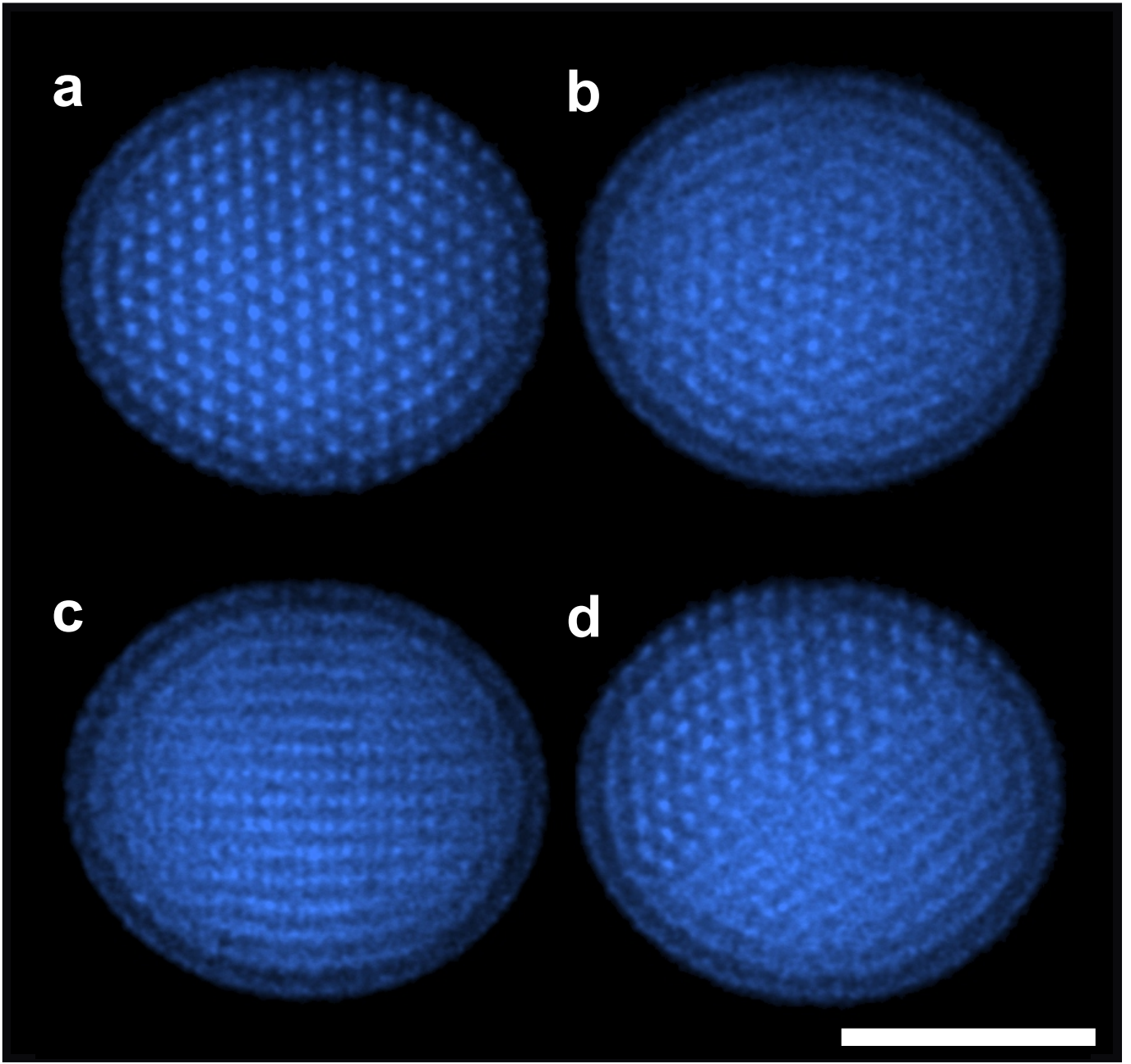

Drewsen, Fig. 4